\documentclass[journal]{IEEEtran}
\ifCLASSINFOpdf
   \usepackage[pdftex]{graphicx}
   \graphicspath{{../pdf/}{../jpeg/}}
   \DeclareGraphicsExtensions{.pdf,.jpeg,.png}
\else
   \usepackage[dvips]{graphicx}
   \graphicspath{{../eps/}}
   \DeclareGraphicsExtensions{.eps}
\fi

  \usepackage[caption=false,font=footnotesize]{subfig}
\usepackage{cite}

\begin{document}

\title{Physical Encryption and Pilot Direction Reversal With Channel Modulation Techniques}

\author{Gerald~Artner,~\IEEEmembership{Member,~IEEE}
\thanks{G. Artner is with the T\"UV AUSTRIA Group, Deutschstra\ss e 10, 1230 Vienna, Austria; formerly with with the Institute with Telecommunications, Technische Universit\"at Wien, Gu\ss hausstra\ss e 25, 1040 Vienna, Austria. e-mail: gerald.artner@nt.tuwien.ac.at, website: geraldartner.com}
}

\markboth{Journal of \LaTeX\ Class Files,~Vol.~xx, No.~x, Month~year}%
{Gerald Artner: Physical Encryption and Pilot Direction Reversal With Channel Modulation Techniques}

\maketitle

\begin{abstract}
Encryption for channel modulation techniques on a physical basis is considered.
A channel modulation scheme is proposed in which the selected channel depends on both transmitter and receiver configuration.
The planned selections of the receiver become the encryption key.
The communication system is prototyped in an anechoic chamber, where channels are distinguished based on phase differences.
It is explored that, as the data is transmitted solely through the selection of the channel, the receiver could transmit the pilot signals and the transmitter could remain passive, i.e. the transmitter does not transmit electromagnetic waves.
\end{abstract}

\begin{IEEEkeywords}
channel, encryption, modulation, spatial
\end{IEEEkeywords}

\IEEEpeerreviewmaketitle

\section{Introduction}
\IEEEPARstart{C}{hannel} modulation techniques have recently been proposed to transmit information through the selection of wireless communication channels \cite{mesleh2008,jeganathan2009,renzo2011,yildirim2017,mokh2018}.
This can be done in contrast or in addition to traditional communication schemes where information is modulated onto the transmit signal.

\section{Considerations on Encryption with Channel Modulation Techniques on a Physical Basis}

\begin{figure}[!t]
\centering 
\subfloat[]{\includegraphics[width=0.17\textwidth]{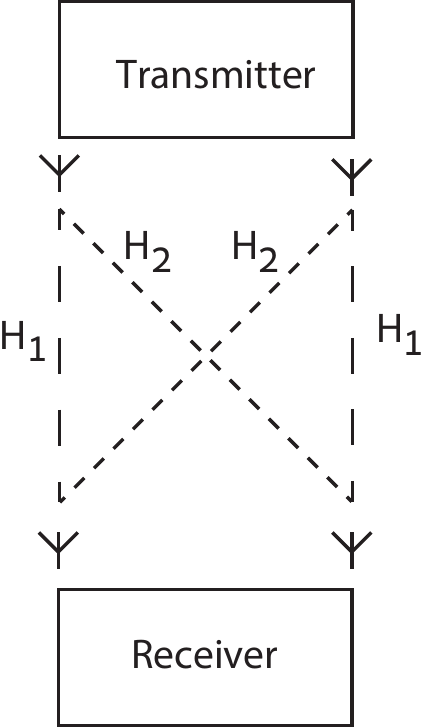}\label{fig_scheme_theory}} \hfill
\subfloat[]{\includegraphics[width=0.17\textwidth]{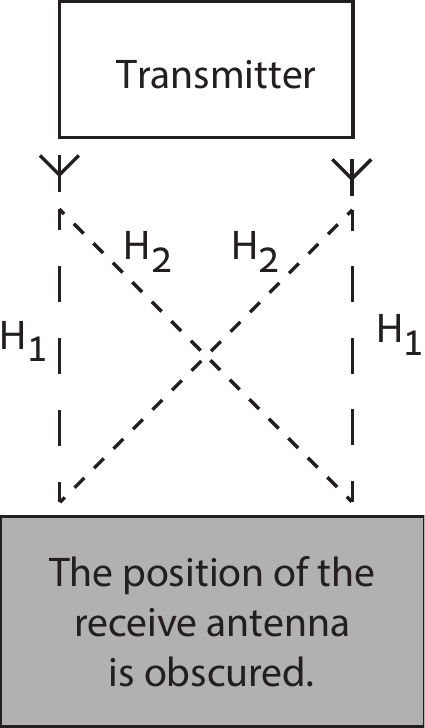}\label{fig_scheme_theory_obscured}} 
\caption{a) A transmitter and a receiver can jointly select a communication channel. b) An outside observer can not determine the selected channel, if the position of the receive antenna is obscured.}
\end{figure}

The wireless communication system in Fig.~\ref{fig_scheme_theory} is considered.
The transmitter can choose between different ways to transmit, which is sketched as two antenna positions in Fig.~\ref{fig_scheme_theory}.
The receiver can also choose between different ways to receive, which is again sketched as two antenna positions.
The considered channel behaves such that a channel $H_1$ is selected when the same antenna positions are chosen and a channel $H_2$ is selected when alternating antenna positions are chosen.
The transmitter and the receiver thus jointly select a communication channel.
Information is transmitted by selecting either $H_1$ or $H_2$.
For transmission, the planned selections of the receiver must be known to the transmitter.

The system is considered again in Fig.~\ref{fig_scheme_theory_obscured}, but the position of the receive antenna is obscured.
An outside observer can not determine the selected channel based on the position of the transmit antenna alone.
The (planned) positions of the receive antenna become the encryption key of the communication.

Pairs of transmitter-receiver choices that result in the same channels can not be used to transmit information with channel modulation schemes, however transmitter-receiver pair that results in the same channel as other transmitter-receiver configurations could be used for encryption --- without loss of capacity for channel modulation.

\section{Experiment}

An experiment is performed with the Vienna MIMO Testbed \cite{Lerch2014} in a shielded and anechoic chamber to demonstrate the feasibility of the theoretical considerations.
The anechoic chamber is a harsh environment for channel modulation techniques.
Multi-path propagation is heavily attenuated and practically non-existent.
As a result, channels can not be modulated based on angle-of-arrival or delayed paths.
The amplitudes of the received signals vary only slightly between different antenna positions as they are a result of free space path loss, but not small scale fading.
The phase difference between antenna positions remains as viable quantity to distinguish wireless communication channels.
The limitations imposed by the anechoic environment severely limit the performance of channel modulation techniques, but they make experiments ostensive and reproducible.

\begin{figure}[!t]
\centering
\includegraphics[width=0.35\textwidth]{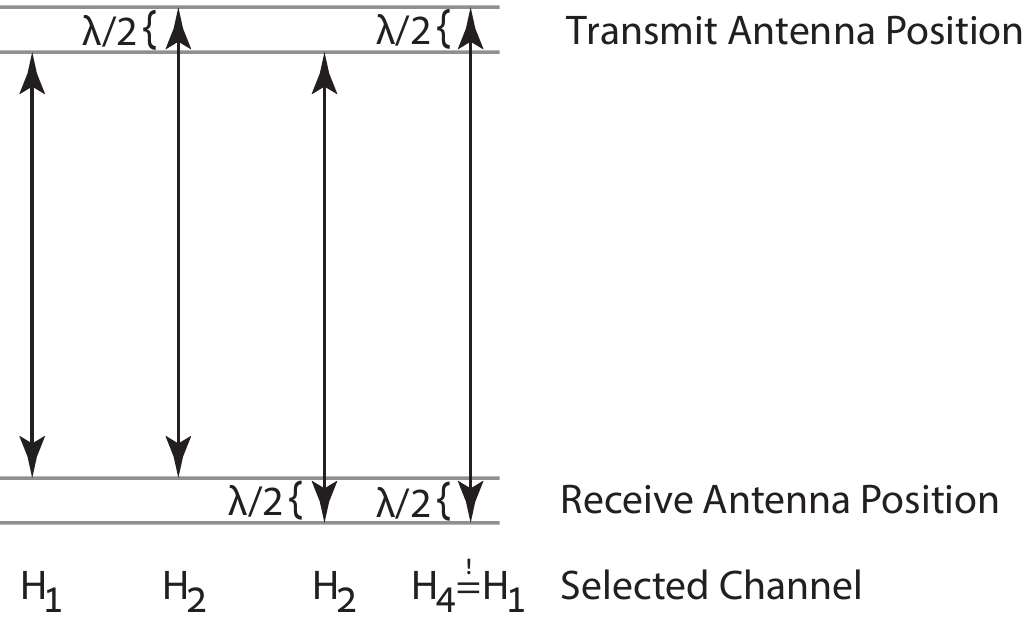}
\caption{The different channels are realized by phase shifts at different antenna positions. Antennas are moved by $\lambda / 2$ which results in additional phase shifts of $\pi$.}
\label{fig_scheme_measurement}
\end{figure}

For the experiment, the transmitter and the receiver are each equipped with one antenna that physically moves between two positions.
The antennas are moved by computerized numerical control (CNC) linear movement units with a high precision of $0.02$\,mm ($0.00016$\,$\lambda$).
The movement units are aligned in a line such that the antennas move together and apart.
The antennas move by a distance of $\lambda / 2$, which results in additional phase shifts of $\pi$, see Fig.~\ref{fig_scheme_measurement}.
$H_4$ is $\lambda$ longer than $H_1$, which results in an additional phase shift of $2 \pi$.
$H_4$ is made equal to $H_1$ by wrapping the phase at $2 \pi$.

\begin{figure}[!t]
\centering
\subfloat[]{\includegraphics[width=0.49\textwidth]{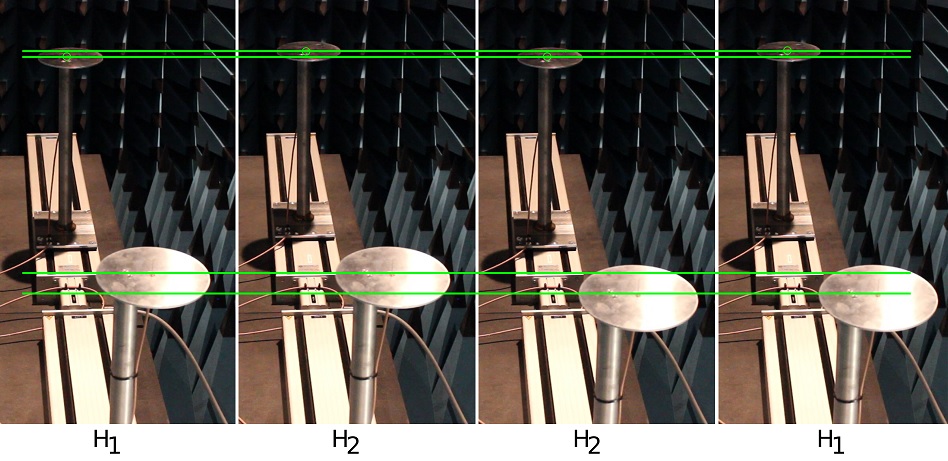}\label{fig_photo_measurement}}\\
\subfloat[]{\includegraphics[width=0.49\textwidth]{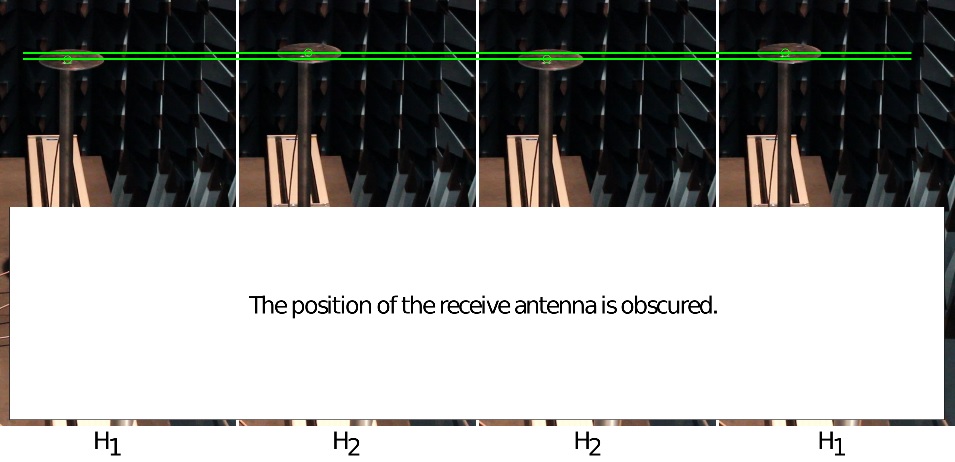}\label{fig_photo_measurement_hidden}}
\caption{Data are transmitted by selecting channels. A channel is realized that is distinguishable by a $\pi$ phase shift at different antenna positions. a) Photos of different antenna positions and their corresponding channels in the experiment. b) The position of the receive antenna is obscured. An outside observer can not determine the selected channel from the position of the transmit antenna alone.}
\label{fig_measurement}
\end{figure}

Both antennas are quarter-wavelength monopole wire antennas for the $2.45$\,GHz industrial, scientific and medical (ISM) frequency band.
The antennas are placed on circular aluminum ground planes with a diameter of $18$\,cm that are elevated from the movement units by $51$\,cm tall posts.
Both antennas are connected to a vector network analyzer (VNA) outside the chamber with coaxial cables.
The ports were through, open, short, match (TOSM) calibrated up to the antenna connectors.
The VNA measures the scattering parameters and makes them available to both transmitter and receiver without added assumptions on accuracy, quantization or delay.
The phases at the antenna positions are measured during the initializing phase and channels are later discriminated by hard decisions in the middle of the initially measured phases.
The amplitude data of the measured scattering parameters are discarded.
The linear movement units and the VNA are controlled by a MATLAB script on a laptop computer, which also acts as both transmitter and receiver.
This produces intuitive visual representations of the channel modulation and encryption properties, see Fig.~\ref{fig_photo_measurement}.
A video is available in the online version.

Plain text messages were transmitted with the described channel modulation technique.
The characters were converted into a bit stream via the American Standard Code for Information Interchange (ASCII) and the bit stream was mapped onto channels $H_1$ and $H_2$.
For each bit the channels were realized by moving the antennas to their positions.
Receive antenna positions were generated as pseudo random sequence and made available to the transmitter, which choose the transmit antenna positions accordingly.
The transmitter's position is determined as message XOR encryption key.
Once the antennas arrived at their positions, the VNA measured the scattering parameters and calculated the phase.
The phase measurements were the channels $\widehat{H}_1$ and $\widehat{H}_2$ estimates, which are then mapped on the received bit stream and converted to the received ASCII characters.
An outside observer would be able to decode the transmitted information.
The antenna positions were visible and the mapping from the channels to the bitstream and the ASCII encoding could be found by trial and error.

\begin{figure}[!t]
\centering
\includegraphics[width=0.49\textwidth]{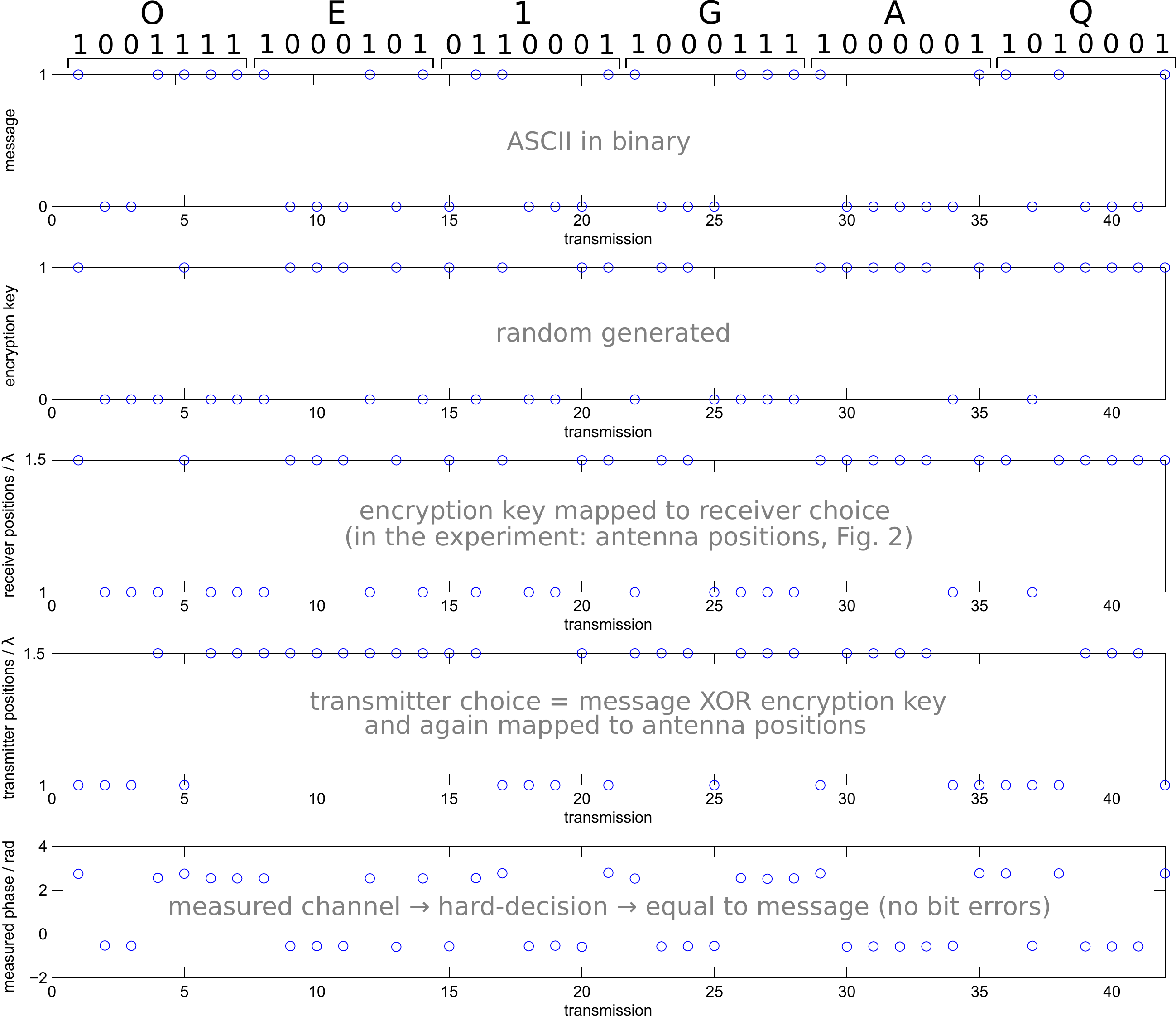}
\caption{Different signals and the measured channel of a sample transmission: the callsign OE1GAQ.}
\label{fig_signals}
\end{figure}

The described experiment was used to establish two-way communication between amateur radio stations OE1GAQ and OE1XTU (operator OE1VMC) on 2018-12-12, 15:50 UTC.
The experiment was repeated on 2018-12-13, 11:35 UTC in a two-way contact between the stations OE1GAQ and OE1ABU.
The recorded signals and the measured channels of a sample transmission are given in Fig.~\ref{fig_signals}.
The steps after measuring the channel are omitted in Fig.~\ref{fig_signals}, because it is already obvious from the measured phase, that the message was successfully received.
No bit errors were experienced during transmission.

The experiment is considered again in Fig.~\ref{fig_photo_measurement_hidden}, but the positions of the receive antenna are obscured.
An outside observer can not determine the selected channel based on the position of the transmit antenna alone.
Some practical considerations on obscuring channel selection follow.
Theoretically, the transmit antenna position might be obscured as well, but the phase shift of the transmitted signal can be measured.
Visually hiding antennas is already state-of-the-art with radar domes.
In a further step, the receiver might be equipped with multiple antennas that are either connected to the receiver or terminated by a matched load.
The receiving station might also receive the signals from both antennas and simply discard the signal from the unselected antenna in software.
In a more general view, the receiver selects communication channels, which can be done with technical solutions other than changing antenna position, e.g. pattern reconfigurable antennas, beamsteering, frequency hopping, selecting delayed taps, smart reflecting surfaces etc.

\section{Considerations on the Direction Reversal of Pilot Signals with Channel Modulation}

Channel modulation schemes have another peculiar property.
Theoretically, because the transmission of information only depends on the selected channel, it is completed once both transmitter and receiver have selected an antenna.
Practically, the receiver could therefore be the station that broadcasts the channel sounding pilot signal.
This concept was successfully tested in the proposed laboratory experiment.
\begin{enumerate}
	\item The receiver selects a ``receive'' antenna and broadcasts the pilot signal.
	\item The transmitter knows the (planned) receive antenna location and selects a channel with its ``transmit'' antenna position.
	\item When the pilot signal is received, this immediately determines the selected channel.
	The transmitter is still the station that transmits the information, although it is the station that receives the pilot signal.
	\item In the experiment, the selected channel is immediately available to the receiver via the channel measurement from the VNA.
\end{enumerate}
The proposed reversal was performed by trivially switching the directions of the measured scattering parameters from $S_{21}$ to $S_{12}$.
The VNA acts as a sort of additional communication channel, that feeds information about the selected radio channel back to the receiver.
The existence of the ``VNA channel'' of course makes the original communication over the radio channels obsolete.
Nevertheless, it's a fun little thought experiment to perform.

In a real-world environment the selected channel is still determined the instant when the transmitter receives the channel sounding signal, but the receiver does not have omniscient channel knowledge via a VNA.
However, having covert channels that feed back the selected channel might be desirable in some practical scenarios to realize \emph{seemingly} passive transmitters.

\section{Conclusion}

Encryption is demonstrated for channel modulation techniques where the receiver can influence the selected channel.
No alterations to channel modulation or the system are performed to encrypt the communication.
Encryption is achieved by simply obscuring which channel is selected by the receiver.
A simple receiver with two antennas can already fulfill this requirement.
Eavesdropping is further physically impeded, as transmitter choices that are distinguishable at the receiver are not necessarily distinguishable at a different location.
Any transmitter-receiver configurations that results in the same channel as other transmitter-receiver configurations could be used for encryption --- without loss of capacity for channel modulation.
It appears that encryption is an intrinsic possibility with channel modulation schemes.

Considerations suggest that the transmitter-to-receiver direction of the pilot signal can be reversed to a receiver-to-transmitter pilot signal in channel modulation techniques.
This has been tested in a laboratory environment.
However, it might not be possible to feed information about the selected channel back to the receiver without additional resources.

\section*{Acknowledgment}

\noindent
The author thanks C.F.~Mecklenbr\"auker, callsign OE1VMC, and S.~Pratschner, callsign OE1ABU, both of Radio-Amateur-Klub der TU Wien, callsign OE1XTU, for testing the proposed method in two-way contacts.
QSL.

\section*{Conflict of Interest}

\noindent
The author has filed a patent application \cite{patent}.

\ifCLASSOPTIONcaptionsoff
  \newpage
\fi

\vfill 


\begin{thebibliography}{1}


\bibitem{mesleh2008}
R.Y.~Mesleh, H.~Haas, S.~Sinanovi\'c, C.W.~Ahn, and S.~Yun, ``Spatial modulation,'' \emph{IEEE Transactions on Vehicular Technology}, vol. 57, no. 4, pp. 2228-2241, 2008.
\bibitem{jeganathan2009}
J.~Jeganathan, A.~Ghrayeb, L.~Szczecisnki, and A.~Ceron, ``Space Shift Keying Modulation for MIMO Channels,'' \emph{IEEE Transactions on Wireless Communications}, vol. 8, no. 7, pp. 3692-3703, 2009.
\bibitem{renzo2011}
M.~Di~Renzo, H.~Haas, and P.M.~Grant, ``Spatial modulation for multiple-antenna wireless systems: a survey,'' \emph{IEEE Communications Magazine}, vol. 49, no. 12, pp. 182-191, 2011.
\bibitem{yildirim2017}
I.~Yildirim, E.~Basar, and I.~Altunbas, ``Quadrature channel modulation,'' \emph{IEEE Wireless Communications Letters}, vol. 6, no. 6, pp. 790-793, 2017.
\bibitem{mokh2018}
A.~Mokh, M.~Crussi\'ere, M.~H\'elard, and M.~Di~Renzo, ``Theoretical Performance of Coherent and Incoherent Detection for Zero-Forcing Receive Antenna Shift Keying,'' \emph{IEEE Access}, vol. 6, pp. 39907-39916, 2018.
\bibitem{Lerch2014}
M.~Lerch, S.~Caban, M.~Mayer, and M.~Rupp, ``The Vienna MIMO Testbed: Evaluation of Future Mobile Communication Techniques,'' \emph{Intel Techn. J.}, vol. 18, no. 3, pp. 58-69, 2014.
\bibitem{patent}
G.~Artner, ``Verschl\"usselung durch Funkkanalmodulation,''Austrian Patent Applicaton A21/2020.

\end{thebibliography}
\end{document}